\documentclass[10pt,journal,final,twocolumn]{IEEEtran}
\IEEEoverridecommandlockouts
\usepackage{cite}
\usepackage{amsmath,amssymb,amsfonts}
\usepackage{algorithmic}
\usepackage{graphicx}
\usepackage{textcomp}
\usepackage{xcolor}
\usepackage{float}
\usepackage{caption}
\usepackage{subfigure}
\usepackage{amsmath}
\usepackage{epstopdf}
\usepackage[ruled]{algorithm2e}
\usepackage{hyperref}
\allowdisplaybreaks[4]
\hypersetup{hypertex=true,
            colorlinks=true,
            linkcolor=blue,
            anchorcolor=blue,
            citecolor=blue}
\makeatletter
\renewcommand{\maketag@@@}[1]{\hbox{\m@th\normalsize\normalfont#1}}
\makeatother

\def\BibTeX{{\rm B\kern-.05em{\sc i\kern-.025em b}\kern-.08em
    T\kern-.1667em\lower.7ex\hbox{E}\kern-.125emX}}

\begin{document}

\title{Sensing, Communication, and Control Co-design for Energy Efficient Satellite-UAV Networks}
\author{Tianhao Liang, \IEEEmembership{Graduate Student Member,~IEEE}, Huahao Ding, Yuqi Ping, Bin Cao, \IEEEmembership{Member,~IEEE}, Tingting Zhang, \IEEEmembership{Member,~IEEE}, \IEEEmembership{Member,~IEEE}, Qinyu Zhang, \IEEEmembership{Senior Member,~IEEE}

\thanks{
Tianhao Liang, Huahao Ding, Yuqi Ping, Bin Cao, Tingting Zhang and Qinyu Zhang are with School of Electronics and Information Engineering, Harbin Institute of Technology (Shenzhen), Shenzhen, China. Tingting Zhang and Qinyu Zhang are also Pengcheng Laboratory (PCL), Shenzhen, 150001, China (e-mail: liangth@stu.hit.edu.cn; hitszdhh@163.com; hitpingyq@163.com; caobin@hit.edu.cn; zhangtt@hit.edu.cn; zqy@hit.edu.cn).
}

}

\maketitle

\begin{abstract}
Traditional terrestrial communication infrastructures often fail to collect the timely information from Internet of Thing (IoT) devices in remote areas. To address this challenge, we investigate a Satellite-unmanned aerial vehicles (UAV) integrated Non-terrestrial network (NTN), where the UAV is controlled by remote control center via UAV-to-Satellite connections. To maximize the energy efficiency (EE) of the UAV, we optimize the UAV trajectory, power allocation, and state sensing strategies, while guaranteing the control stability and communication reliability. This challenging problem is addressed using an efficient algorithm, incorporating a Deep Q-Network (DQN)-based trajectory determination, a closed form of power allocation, and one-dimensional searching for sensing. Numerical simulations are conducted to validate the effectiveness of our approach. The results showcase the data size of collection has a greater impact than transmission power, and reveal the relationship among sensing interval, communication maximum power and control performance.  This study provides promising solutions and valuable insights for efficient data collection in remote IoT.
\end{abstract}

\section{Introduction}
\subsection{Background and Motivation}
Motivated by the development of wireless technologies in terms of communication capacity, latency, and reliability, the Internet of Things (IoT) has captured significant attentions in recently years. IoT devices have become integral components in various applications, ranging from environmental monitoring and industrial automation. The data collected by these devices is crucial for making informed decisions \cite{WZJIOTJ}.

However, one of the significant challenges is the timely and reliable connection in some critical scenarios, such as deep canyons and remote area, where are always lack of the terrestrial networks due to deployment and maintenance cost \cite{UAVdis}. Those limitations promote the exploration of alternative frameworks that can overcome these connectivity barriers and ensure robust data collection \cite{lthicc}.

In this context, non-terrestrial networks (NTNs), which include unmanned aerial vehicles (UAVs), high altitude platforms (HAPs), low Earth orbit (LEO) satellites,  medium Earth orbit (MEO) satellites and geostationary Earth orbit (GEO) satellites networks, emerge as a promising solution to address above challenges. In our pervious work \cite{LTHNETWORK}, we pointed a Satellite-UAV framework to support the seamless localization and communication. The Satellite-UAV framework leverages the strengths of both satellite and UAV technologies to create a flexible and scalable platform for data collection. Satellites offer wide coverage and communication links to remote ground core network, while UAVs offer mobility and the ability to operate at lower altitudes, closer to ground IoT devices (GDs) \cite{LJTWC}.

The design and implementation of a Satellite-UAV framework  involve addressing several key challenges related to sensing, communication, and control. To complete the data collection of a set of GDs, the state of UAVs should execute state sensing periodically or aperiodically and transmit to the remote center for its control commends determination \cite{2019ac}. The importance of sensing for the UAV lies on two aspects. On the one hand, the localization of UAV, obtained by Global Navigation Satellite System (GNSS), is crucial for the random access in the 3rd generation partnership prohect (3GPP) Rel-17 NTN standard. However, the GNSS can easily be disturb by some GNSS spoofing technology. The jointly GDs and GNSS sensing technology can effective solve this problem. On the other hand, the vertical localization accuracy of the UAV can only approach over 3m undergoing high dynamic motions \cite{TVT2023} \cite{UAV_localization}. The state error will impair control performance, even lead to the instability of this remote control system. 

Another critical consideration UAVs-involved work is the energy consumption \cite{ZYTWC}. UAVs are often constrained by limited battery life, which can significantly impact their operational range and duration. Frequent sensing can reduce the control cost and improve control performance,  but it will lead to high energy consumption. Therefore, to complete the data collection, how to design the sensing, communication, and control strategies for a energy efficient Satellite-UAV network remain a challenge \cite{Tcom2021}.

\subsection{Related Work}
Current investigations are mainly focused on the communication and control co-design issue, which can primarily be categorized into two kinds. The first kind considers either communication or control performance as the optimization objective while considering the other as constraints. In the context of control performance optimization, Eisen et al. minimize control cost subject to strict latency requirements by dynamically adjusting the packet delivery rate targets of devices, selectively scheduling and utilizing missions \cite{Lowlatencyco}. In \cite{7798258}, Gatsis et al. formulate a dynamic sensor transmission policy for wireless control systems, optimizing the control cost under the constraint of average communication resources. In contrast, some studies optimizes communication performance under control constraints, such as, Gatsis et al. formulates the optimal design of channel-aware scheduling and power allocation to minimize total communication power consumption while meeting control performance requirements for all systems \cite{7070754}. Chang et al. solve the resource allocation problem of ultra-reliable and low-latency communication (URLLC) in real-time wireless control systems based on uplink transmission by optimizing bandwidth and transmission power allocation in URLLC and controlling the convergence speed under communication and control constraints \cite{Changtvt}. The second kind combines both communication and control factors into the optimization objective simultaneously. In \cite{AOIIOT}, Wang et al. minimize both control cost and communication energy consumption under the consideration of the average Age of Information impact in co-design.

When wireless networks are constrained, such as in military environments, wireless scheduling in NTN scenario is the effective manner. Han et al. employed the NTN to optimize the trajectory of UAV to accomplish combat missions in adversarial interference environments by reinforcement learning methods \cite{Chenh}. Wang et al. iteratively optimize smart device connection scheduling, power control, and UAV trajectory design to achieve maximum system capacity \cite{WYSJ}. Wei et al. optimize the average offloading time of communication using a multi-agent Q-learning algorithm, under the constraints of control performance \cite{Pengw}. In addition to the communication control co-design framework, Lei et al. consider the integration of UAV perception in Integrated Satellite-UAV Networks control scenarios \cite{WCL2023}. Liu et al. address the uplink transmission communication issue by jointly optimizing UAV three-dimensional trajectory design, UAV-sensors association, and sensors' transmission power. Their approach aims to maximize long-term network capacity while simultaneously minimizing the sensors' total energy consumption using DRL \cite{Liuj}.

Overall, most investigations only focus on the communication control co-design for optimizing communication or control metrics. In NTN scenarios, many studies assumed the perfect control of UAVs while designing their trajectory.  Moreover, studies in this context rarely consider the sensing issue, which is essential for communication and control \cite{JSAC}.

Motivated by above issues, we aspire to design the sensing scheduling, power allocation as well as UAV control strategies in a UAV-aided data collection system.  The energy efficiency (EE) is maximized while maintaining control stability and communication reliability.  The main contributions of this work are outlined as follows.
\begin{itemize}
  \item We first establish a framework for UAV trajectory tracking, data uploading, and state sensing in the UAV-aided data collection, where linear quadratic regulator (LQR) is utilized to control UAV, and UAV-Satellite uplink is used for data uploading. The state sensing of UAV is performed by GNSS and GD's measurements jointly. 
       To the best of our knowledge, such a system has not been explored in current literatures.
  \item Subsequently, we formulate a co-design optimization problem aiming to maximize EE of the UAV by optimizing its trajectory, uploading power, and state sensing strategies,  while guaranteing the control stability and date transmission reliability.  To tackle this complicated problem, we propose an efficient algorithm to obtain a sub-optimal solution, where Deep Q-Network (DQN)-based trajectory determination and closed form of power allocation are provided.  
  \item  Lastly, we conduct the extensive simulations to demonstrate the effectiveness of the proposed framework. The result showcase the great EE performance and reveal the relationship of sensing, communication, and control in the data collection task. 
\end{itemize}

The rest of this paper is organized as follows. We introduce the detailed system model for communication, control, and sensing, respectively, in section II. Section III presents an energy efficient problem formulation. In section IV, we propose an efficient algorithm to solve this optimization problem. The simulation results are provided in section V. Finally, Section VI draws the conclusion of this paper. 

{\it Notations}: We use lowercase and uppercase bold-symbols, calligraphic uppercase characters to denote vectors, matrices and sets, respectively. The superscript $(\cdot)^\text{T}$ and $\|\cdot\|$ denote the transpose and Euclidean norm of its argument, respectively.

\begin{figure}[h]
\centering
\includegraphics[width=1\columnwidth]{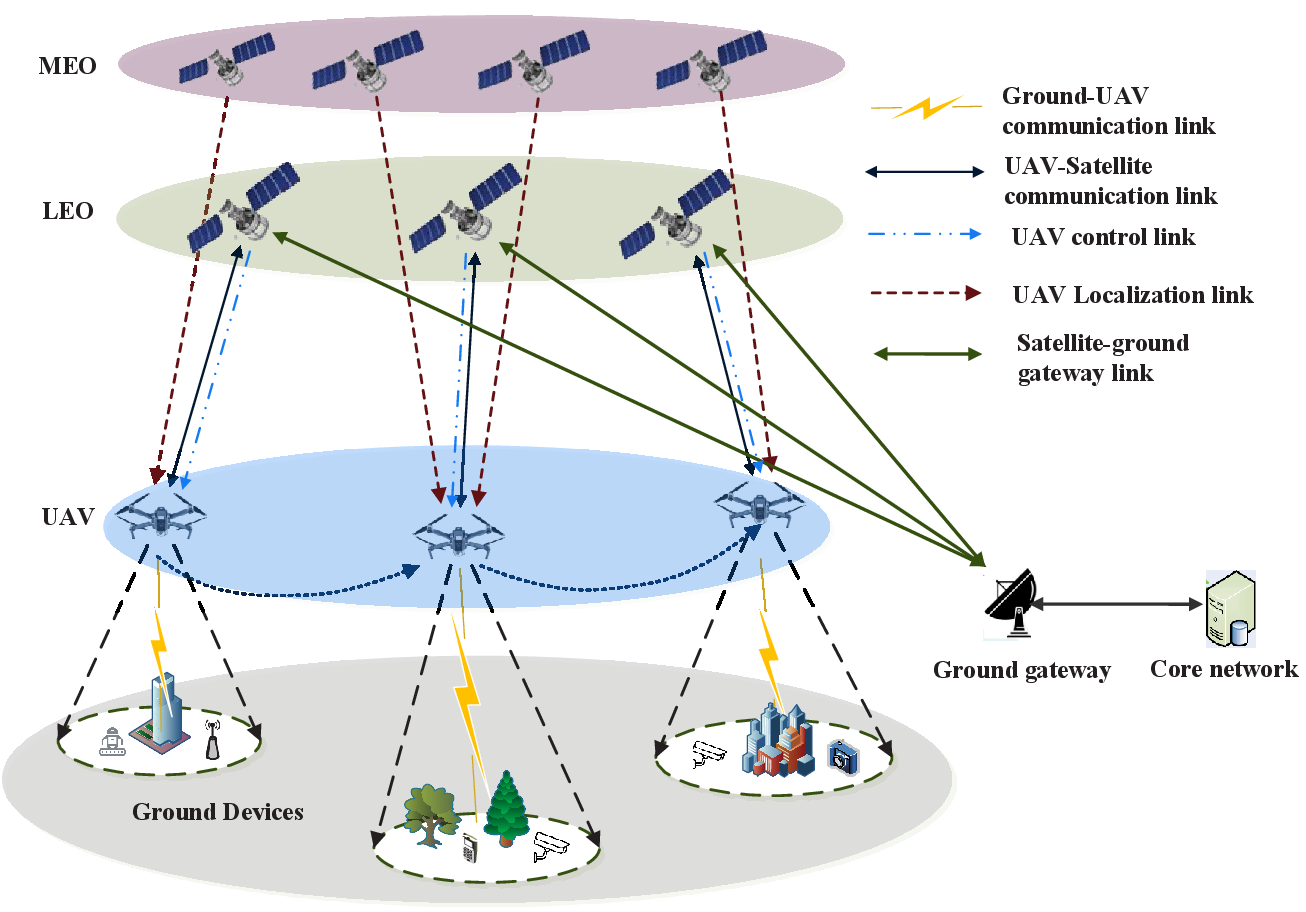}
\caption{The scenario of remote UAV control for data collection.}
\label{scenario}
\end{figure}
\begin{figure}[h]
\centering
\includegraphics[width=1\columnwidth]{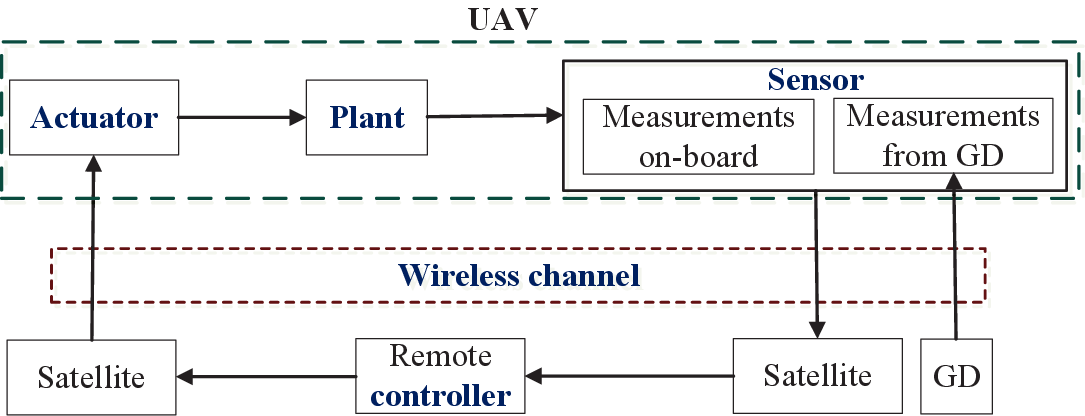}
\caption{The structure of remote UAV control.}
\label{structure}
\end{figure}

\section{System Model}
\label{sec:II}
In Fig. \ref{scenario}, we illustrate the architecture of Satellite-UAV NTN,  comprising satellites for communication and UAVs for relay information. A set $\mathcal{N}_\text{d}$ of $N_\text{d}$ ground devices (GDs) is deployed in advance within the area to perform specific tasks, such as environmental monitoring. Without loss of generality, we assume there is one UAV in this system to execute the data collection task. Specifically, the UAV is controlled by the LEO, which receives control commends from remote ground control center to track the predetermined trajectory, and forwards the collected data from UAV to the ground gateway for processing. 

The process of the proposed data collection task under Satellite-UAV structure can be brief described as three steps. Firstly, the remote control center should given a reference trajectory to perform the data collection for target GDs. Secondly, the UAV is controlled remotely to track this trajectory based on its current state. Finally, the UAV collects the data from GDs and uploads them to Satellite for forwarding to ground gateway.  

As time progresses, the state prediction uncertainty and control instability will become intolerable if the state of the UAV is not sensed and transmitted to the control center. To guarantee the control stability and task completion, the state of UAV needs to be sampled sufficiently and transmitted to the remote control center for calculating appropriate control actions\footnote{The remote control center will provided a set of predictive control commands and transmits to the UAV in each control commands calculation.}. 
Such an system can be describes by a wireless networked control structure in Fig. \ref{structure}, where the actuator, sensor, and control are separated and connected by wireless network.  The actuator is located at the UAV for executing the control commands. In our previous research \cite{lthiotj}, we emphasized the importance of the UAV in enhancing the localization accuracy for uncertainty GDs. Conversely, accurate GDs can also improve the UAV's localization accuracy by executing {\it sensing} operations \cite{LYIOT}.  Therefore, the sensing result of sensor is determined joinitly by on-board GNSS information (from MEO) and  measurements from GDs. The controller is located at the remote cloud, which calculates the control command based on the results from sensor and transmits them to the actuator.

With following the reference trajectory, the UAV is hovering for collecting the data from GDs, and flying for uploading to the satellite. 
Denote ${\bf x}(t)=\left[p_{x}^\text{u}(t),p_{y}^\text{u}(t),p_{z}^\text{u}(t),v_{x}^\text{u}(t),v_{y}^\text{u}(t),v_{z}^\text{u}(t)\right]^\text{T}$ as the state space of the UAV, and let ${\bf x}_{i}^\text{d}=\left[p_{i,x}^\text{d},p_{i,y}^\text{d},p_{i,z}^\text{d},0,0,0\right]^\text{T}$ be the state information of pre-deployed GD $i\in\mathcal{N}_\text{d}$. 



\subsection{Control for Trajectory Tracking}
For UAV control, we model the wireless control system as a linear time-invariant system \cite{WCL2023}. The discrete time linearized control equation for UAV state evolution at time slot $k$ can be given by 
\begin{equation}\label{cm}
{\bf x}(k+1)={\bf A}{\bf x}(k)+{\bf B}{\bf u}(k)+{\bf w}(k),
\end{equation}
where ${\bf A}\in \mathbb{R}^{6\times6}={\bf A}_1\bigotimes{\bf I}_{3\times3}$ and ${\bf B}\in\mathbb{R}^{6\times3}={\bf B}_1\bigotimes{\bf I}_{3\times3}$, are the state transition matrix and control matrix, respectively. The term of $\bigotimes$ is the Kronecker product, and ${\bf A}_1$, ${\bf B}_1$ are expressed as
\begin{equation}
{\bf A}_1=\left[
\begin{array}{cc}
  1 & T_\text{s} \\
  0 & 1
\end{array}\right], \quad
{\bf B}_1=\left[
\begin{array}{c}
  \frac{1}{2}T_\text{s}^2  \\
  T_\text{s}
\end{array}\right],
\end{equation}
where $T_\text{s}$ is the sampling interval.
In this paper, we assume the UAV is controlled by adjusting accelerations, indicating  ${\bf u}(k)=\left[a_{x}(k),a_{y}(k),a_{z}(k)\right]^\text{T}$. The term ${\bf w}_{k} \sim\mathcal{N}({\bf 0}_{6\times1},{\bf R})$ is the state transmission noise. 

As mentioned in \cite{Tcom2021}, if the spectral radius of $\bf A$ is greater than unit, the plant's state will grow infinitely over time unless given a proper control action ${\bf u}(k)$.
For a given state, the controller output can be computed utilizing the LQR to track the reference state, and the optimal action is given by the following feedback control law
\begin{equation}\label{LQR}
{\bf u}(k)=-\underbrace{({\bf B}^\text{T}{\bf P}{\bf B}+\boldsymbol{\epsilon})^{-1}{\bf B}^\text{T}{\bf P}{\bf A}}_{\text{LQR control gain} \!\!\!\!\!\!\quad{\bf K}}\left({\bf x}(k)-{\bf x}_\text{r}(k+1)\right),
\end{equation}
where $\boldsymbol{\epsilon}$ is a positive definite weight matrix of action cost, and ${\bf x}_\text{r}(k+1)$ is realistic expected reference state of next time slot, which is calculated according to a specific objective function and hovering points \cite{LJTWC}. The term ${\bf P}$ is obtained by solving the discrete time algebraic Riccati equation
\begin{equation}
{\bf P}={\bf A}^\text{T}{\bf P}{\bf A}-{\bf A}^\text{T}{\bf P}{\bf B}({\bf B}^\text{T}{\bf P}{\bf B}+\boldsymbol{\epsilon})^{-1}{\bf B}^\text{T}{\bf P}{\bf A}+{\bf Q},
\end{equation}
where ${\bf Q}$ is the positive definite weight matrix of state deviation.

From the equation (\ref{LQR}), we can find that the control action of the UAV is determined by the error between current state and reference state, indicating the higher uncertainty of UAV state will lead to higher control cost.  

\subsection{Communication for Data Collection}
For UAV-to-Satellite communication, the channel gain can be regarded as the following free space path loss model. Due to the considerable distance between the UAV and the LEO, we assume that the service radius of each LEO can be neglected compared with its altitude for the sake of analysis \cite{IOT2019}. Therefore, the channel gain between satellite and UAV can be formulated as 
\begin{align}
g_\text{s}=\frac{g_0}{\left(H_\text{s}-p_z^\text{u}\right)^2}\approx\frac{g_0}{H_\text{s}^2},
\end{align}
where $g_0$ is the channel power gain at the reference distance. Then, the data rate can be calculated by 
\begin{align}
R_\text{s}(k)=B_{\text s}\log_{2}\left({1+\frac{p(k)g_\text{s}}{\sigma_0^2}}\right),
\end{align}
where $B_\text{s}$ is the bandwidth for UAV-to-Satellite uplink. The term $p(k)$ is the uplink power of the UAV, and $\sigma_0^2$ denotes the variance of the additive noise power at the receiver.

For Ground-to-UAV communication, we utilize the model in \cite{WCL2014} due to its simplicity and generality.
The connection link between the UAV and GD $n$ at time slot $k$ can be described as line-of-sight (LoS) and non-line-of-sight (NLoS) with probabilities ${P}_{n}^{\text{LoS}}(k)$ and ${P}_{n}^{\text{NLoS}}=1-{P}_{n}^{\text{LoS}}(k)$, respectively, in which
\begin{align}\label{PLOS}
{P}_{n}^{\text{LoS}}(k)=\frac{1}{1+a\exp(-b(\varphi_{n}(k)-a))},
\end{align}
where $a$ and $b$ are environment parameters, $\varphi_{n}(k)=\arctan{\left(\frac{p_{z}(k)}{d_{n}^h(k)}\right)}$ is the elevation with the horizontal distance between GD $n$ and the UAV as $d_{n}^h(k)=\sqrt{(p_{n,x}^\text{d}-p_{x}^\text{u}(k))^2+(p_{n,y}^\text{d}-p_{y}^\text{u}(k))^2}$. 
Then, the average path loss is computed by
\begin{align}\label{avPL}
\overline{L}_{n}(k)={P}_{n}^{\text{LoS}}(k)L_\text{LoS}(k)+P_{n}^{\text{NLoS}}(k)L_\text{NLoS}(k),
\end{align}
where the path loss of LoS link and NLoS link are expressed by  $L_\text{LoS}(k)=\left(\frac{4\pi f_c d_{n}(k)}{c}\right)^2\eta_\text{LoS}$ and  $L_\text{NLoS}(k)=\left(\frac{4\pi f_c d_n(k)}{c}\right)^2\eta_\text{NLoS}$, respectively. The term $f_c$ and $c$ are the carrier frequency and speed of light, and $d_{n}(k)=\sqrt{d_{n}^h(k)^2+\left(p_{z}^\text{u}(k)-p_{n,z}^\text{d}(k)\right)^2}$ is the relative distance between the UAV and GD $n$, $\eta_\text{LoS}$ and $\eta_\text{NLoS}$ are the excessive path loss of LoS and NLoS, respectively.
The signal to noise ratio (SNR) can be computed by $\text{SNR}_n(k)=\frac{g_rP_{n}(k)}{\overline{L}_{n}(k)\sigma_0^2}$, where $g_r$ is the receiver antenna gain at the UAV, $P_{n}(k)$ is the transmitted power of GD $n$. 
Then, the uplink achievable data rate for GD $n$ can be expressed by
\begin{align}\label{DR}
R_{n}(k)=B_0\log_{2}\left({1+\text{SNR}_n(k)}\right).
\end{align}
where $B_0$ is the bandwidth of Ground-to-UAV link. 
In this paper, we adopt the time division multiple access (TDMA) technology for Ground-to-UAV communication.


\subsection{Sensing for Control Stability}
Typically, the stability is crucial for a control system. For the remote wireless control system, the stability is guaranteed when the state of the UAV satisfies
\begin{align}
\limsup\limits_{T\to \infty}\frac{1}{T}\mathbb{E}\left[\sum_{t=1}^{T}\|{\bf e}(t)\|^2\right]<\infty,
\end{align}
where ${\bf {e}}(t)={\bf{x}}(t) - {\bf x}_\text{r}(t)$ is the trajectory tracking error. 
This condition implies the control stability is up to average energy of the UAV over time. 

\begin{figure}
\centering
\subfigure[]{\includegraphics[width=0.8\columnwidth]{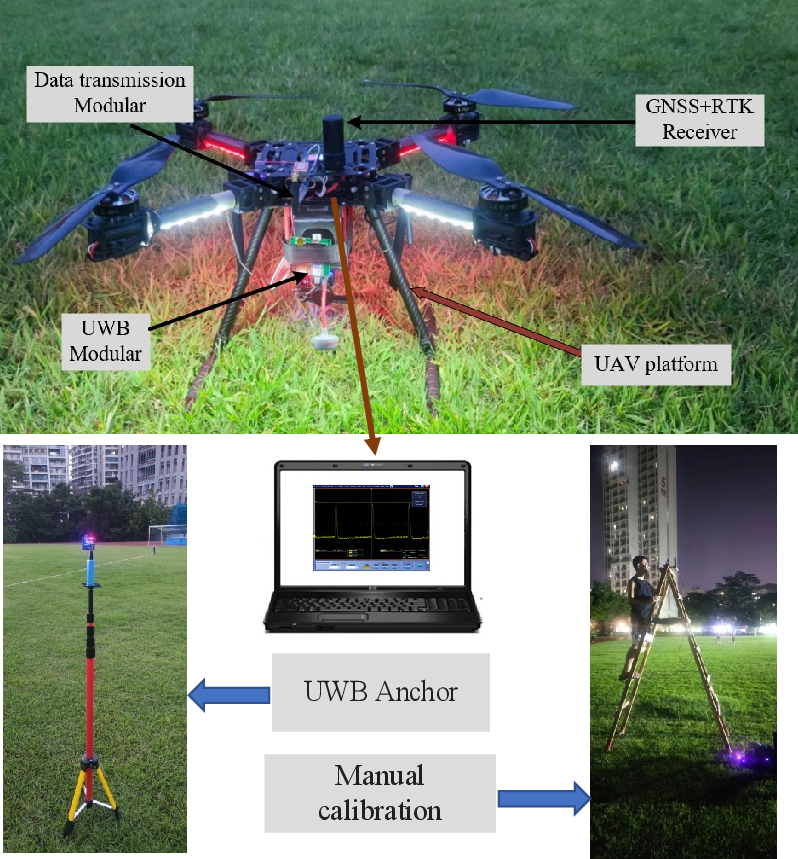}}\\\subfigure[]{\label{localization}\includegraphics[width=0.8\columnwidth]{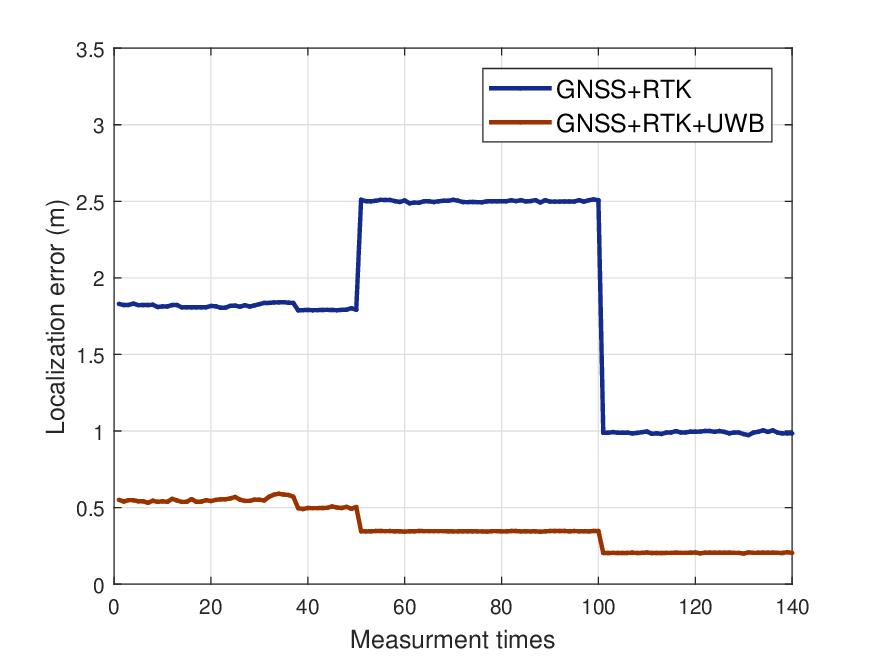}}
\setlength{\abovecaptionskip}{0pt}
\caption{The experiment scenarion and result of UAV localization.}
\label{experiment}
\end{figure}

To guarantee control stability, the UAV needs to transmit sufficient accurate state information to the remote center for calculating control commands. The accurate state information needs to a frequent sensing operations.
However, the increased sensing time results in reduced communication efficiency and higher power consumption.

For UAV state sensing, we utilize the assistance of GD's measurement combination with GNSS. 
In Fig. \ref{experiment}, we carry out an experiment to verify the effectiveness for UAV localization with the assistant of GD. In this experiment,  the UAV is equipped with ultra-wideband (UWB), GNSS and real time kinematic (RTK) receivers. There are four UWB transmitter deployed in ground as the anchors. Under this setting, the GD assistant method can achieve over 70\% accuracy gain than GNSS+RTK method.

In fact, the remote control center may not receive the state information at each time instant due to the sensing scheduling and packet dropouts.  Denote $\gamma(k)$ as a binary sensing scheduling variable. $\gamma(k)=1$ means the UAV performs sensing operation, by receiving the GNSS signals from MEO satellites and pilot signals from GDs at time slot $k$, otherwise, $\gamma_{n}(k)=0$. Then, those measurement are transmitted to the remote controller for fusion and used to compute control commends. 

Differently, the transmission delay of this Satellite-UAV network is much greater than the terrestrial network. Here, we consider the maximum propagation delay of our architecture, where the LEO operates as the transparent payload based on 3GPP R17 NTN standard, indicating that the satellite transmits UAV's state information to the remote ground control center immediately once it receives the imformation.  The maximum propagation delay for UAV-to-Satellite can be calculated by \cite{LEO2019}
\begin{align}
\tau_\text{max}=\frac{\left(R_0+H_\text{s}\right)\sin\Theta_\text{max}}{c\cos\varphi_\text{min}},
\end{align}
where $R_0$ is the Earth radius, $\Theta_\text{max}$ and $\varphi_\text{min}$ are the maximum elevation angle and central angle of the LEO satellite.

Consequently, the state estimation of the UAV in remote control center can be expressed as
 \begin{equation}\label{est}
{\bf x}_{c}(k+\Delta)=\gamma(k)\underbrace{\hat{\bf x}(k+\Delta)}_\text{estimation}+\left(1-\gamma(k)\right)\underbrace{\tilde{\bf x}(k+\Delta)}_\text{prediction},
 \end{equation}
where $\Delta=\lfloor{\frac{2\tau_\text{max}}{\delta_s}}\rfloor$ is the transmission delay in terms of sampling interval $\delta_s$, and $\hat{\bf x}(k+\Delta)$ is the newest state based on estimation in the control center, and $\tilde{\bf x}(k+\Delta)$ is the prediction state. Next, we give two {\it Lemmas} to show the control stability conditions.

{\it Lemma 1\cite{TSP2018}:} If ($\bf A$,$\bf B$) is controllable, under the LQR control law ${\bf u}(k)=-{\bf K}{\bf e}(k)$, where the ${\bf e}_k={\bf x}^\text{b}(k)-{\bf x}_\text{r}(k)$ is remote estimation tracking error covariance.  If $\limsup\limits_{T\rightarrow\infty}\frac{1}{T}\sum_{k=1}^T\mathbb{E}\left[{\bf e}_k\right]<\infty$, the control system is stable.

The {\it Lemma 1} gives the connection between control stability and remote estimation stability. In a word, if the remote estimation is stable, the wireless networked control system is also stable. Subsequently, we introduce {\it Lemma 2} to offer a sufficient condition for guaranteeing the remote estimation stability.

{\it Lemma 2\cite{Tcom2020}:} Denote $\rho(k)=\mathbb{P}(\chi(k)=1)$ as the successful state transmission probability of the UAV at time slot $k$. For $q(k)\geq 2$, the remote estimation is stable at time slot $k+q(k)$ when
\begin{equation}\label{st_con}
\rho(k)>1-\frac{1}{\left(\max\left\{\lambda_m\right\}\right)^{q(k)}}
\end{equation}
where $\lambda_m$ is an eigenvalue of ${\bf A}$, and $q(k)$ is the time duration between current time and next sensing decided time.

{\it Lemma 2} shows the relationship between control stability, sensing scheduling, and sensing successful probability. The successful state transmission probability $\rho(k)=P_\text{G2U}(k)P_\text{U2S}(k)$, where $P_\text{G2U}(k)$ and $P_\text{U2S}(k)$ are the LoS probability on Ground-to-UAV and UAV-to-Satellite links, respectively. Due to the high channel quality of U2S links, the LoS probability between UAV and satellite can be denoted as 1. Consequently, $\rho(k)=\max\left\{{P}_{n}^{\text{LoS}}(k), \forall n\right\}$.

\section{Sensing, Communication and Control Co-design Problem}
\label{sec:III}
The endurance and service performance of UAV system are fundamentally limited by its on-board energy, which is practically finite due to its size and weight constraints \cite{LTHTWC}. Therefore, we aim to optimize energy consumption of the UAV by carefully managing its trajectory, power allocation, and sensing strategies, while maintaining the requisite level of wireless control stability and communication reliability. In the following sections, we provide a detailed calculation of energy consumption at the beginning and then formulate the optimization problem to maximize EE subsequently.  

\subsection{Energy Consumption Calculation}
The energy consumption of the UAV contains sensing, data uploading, and control energy consumptions. The detailed calculation of the energy consumption is presented below.

{\it Energy consumption of control:} The control energy consumption is consisted of the propulsion energy for moving to adjust its trajectory and hovering energy for data collection. The propulsion energy is determined by the velocity and control action in (\ref{cm}) based on given reference state and LQR method.  According to the model proposed in \cite{CL2018}, the control for propulsion energy consumption can be expressed as
\begin{align}\label{CE}
E_\text{f}(k)=\delta\left( \kappa_1\parallel{\boldsymbol\upsilon}(k)\parallel^3+\frac{\kappa_2}{\parallel{\boldsymbol\upsilon}(k)\parallel}\left(1+\frac{\parallel{\bf{u}}(k)\parallel^2}{g^2}\right)\right)
\end{align}
where $\kappa_1$ and $\kappa_2$ are the fixed parameters, and $g$ is the gravitational acceleration. The term  $\delta$ is the length of a time slot. The hovering energy of UAV can be calculated by $E_\text{h}=\delta P_\text{h}$, where $P_\text{h}$ is the hovering power.

{\it Energy consumption of sensing and communication:} The sensing energy is determined by the sensing scheduling strategy. We assume that there is a fixed energy consumption $E_\text{s}$ for each sensing process. The communication energy consumption is determined by the data uploading scheduling from UAV to satellite. For each time slot, it can be calculated by $E_\text{d}(k)=p(k)\delta$.

\subsection{Problem Formulation}
Denoting the total time slot as $T$, the EE of the UAV can be given by
\begin{align}
\eta_\text{EE}=\frac{\sum\limits_{k=1}^TR_\text{s}(k)}{\sum\limits_{k=1}^T\left(E_\text{f}(k)+E_\text{h}(k)+\gamma(k)E_\text{s}+E_\text{d}(k)\right)}.
\end{align}

Our objective is to maximize the EE of the UAV with optimizing its trajectory, power allocation, and sensing strategies,  while maintaining the necessary level of control stability and communication reliability.
Let $\boldsymbol\gamma=\left\{\gamma(k), \forall k\right\}$ and ${\bf P}=\left\{p(k), \forall k\right\}$ denote the variables of communication, sensing scheduling, and power allocation, respectively. Denote ${\bf{X}}=\left\{{\bf{x}}(k), \forall k\right\}$ as the trajcetory of the UAV.

Then, the optimization problem can be formulated as

\begin{align}\label{problem}
\underset{\boldsymbol\gamma,{\bf P},{\bf X}}{\max.}&\quad \eta_\text{EE} \\ \nonumber
 \text{s.t.}\quad  &\text{C1}: \gamma (k)\in\left\{0,1\right\},\forall k\\ \nonumber
 &\text{C2}: {\sum\limits_{k=1}^TR_\text{s}(k)}\geq {\sum\limits_{n=1}^{N_\text{d}}\sum\limits_{k=1}^TR_{n}(k)}, \forall n \\ \nonumber
 &\text{C3}: {\sum\limits_{k=1}^TR_{n}(k)}\geq D_\text{s}, \forall n\\ \nonumber
 &\text{C4}: p(k)\leq p_\text{max},\forall k\\ \nonumber
 &\text{C5}: {\boldsymbol\upsilon}(k)\leq {\boldsymbol\upsilon}_\text{max},\forall k\\  \nonumber
 &\text{C6}: {\bf u}(k)\leq {\bf u}_\text{max},\forall k\\  \nonumber
 &\text{C7}: \rho(k)>1-\frac{1}{\left(\max\left\{\lambda_m\right\}\right)^{q(k)}}, \forall k,
\end{align}
where C1 is the constraint of sensing scheduling,  C2 and C3 are data uploading and collection constraints. C4
represents the constraint of power limitation, where $p_\text{max}$ is the maximum transmitted power of the UAV. Constraints C5 and C6 ensure the safe flight with limited velocity and acceleration. C7 guarantees the control stability. Note that, the objective function is non-convex and the constraints involve both discrete and continuous variables, rendering a mixed-integer nonlinear programming problem. Generally, it is difficult to solve it by using traditional iterative optimization techniques.

\section{Proposed Efficient Algorithm}
In this section, we present an efficient algorithm to find a sub-optimal problem (\ref{problem}), by optimizing  UAV's trajectory , power allocation, and sensing scheduling subsequently. Specifically, we first find the optimal reference trajectory with given hovering points for data collection by a DQN-based method.  Then, we derive the closed-form power allocation results for each data collection processing. Finally, the sensing strategies are provided by a one-dimensional search method to determine the LQR control input.

\subsection{DQN-based UAV Trajectory Determination}
Before conducting perception scheduling, we first need to provide the optimal trajectory for UAV flight between the given hovering points ${\bf x}_\text{r}(i)=\left[p_{x}^\text{r}(i),p_{y}^\text{r}(i),p_{z}^\text{r}(i)\right]^\text{T}$ for collecting the data from the GD $i$, to reach the minimal energy consumption. This issue can be translated into multiple optimal subproblems pertaining to UAV speed optimization. Under the constraints that the initial and final velocities are zero. To simplify the problem and find a reference trajectory ${{\bf{X}}_\text{r}^*}$ that conforms to dynamic constraints, we discretize the UAV's acceleration $\bf{u}$ into distinct values. Then, the optimal problem can be expressed as
\vspace{-2mm}
\begin{align}\label{p2}
\underset{{{\bf{X}}_\text{r}^*},\bf{u}}{\min.}&\quad  E_\text{f}^i \\ \nonumber
 \text{s.t.}\quad &{{\bf{x}}_\text{r}^*}(k_{i,\text{max}}) = {\bf x}_\text{r}(i),\forall i\\ \nonumber
 &{\boldsymbol \upsilon}(k_{i,\text {0}})=0,\forall i \\ \nonumber
 &{\boldsymbol \upsilon}(k_{i,\text {max}})=0,\forall i \\ \nonumber
 &\text{C5},\text{C6}, \text{C7}
\end{align}
where ${{\bf{x}}_\text{r}^*}(k_{i,\text{max}})$ and ${\bf{x}}_\text{r}(i)$ represent the final state of $i$-th trajectory and the hovering state of $i$-th trajectory, $E_\text{f}^i=\sum\limits_{j=1}^{k_{i,\text{max}}}E_\text{f}(j) $ represents the total moving energy used in $i$-th trajectory, $k_{i,\text {max}}$ and $k_{i,\text {0}}$ represent the initial and the final time slot of $i$-th trajectory.

\begin{figure}[h]
\centering
\includegraphics[width=1\columnwidth]{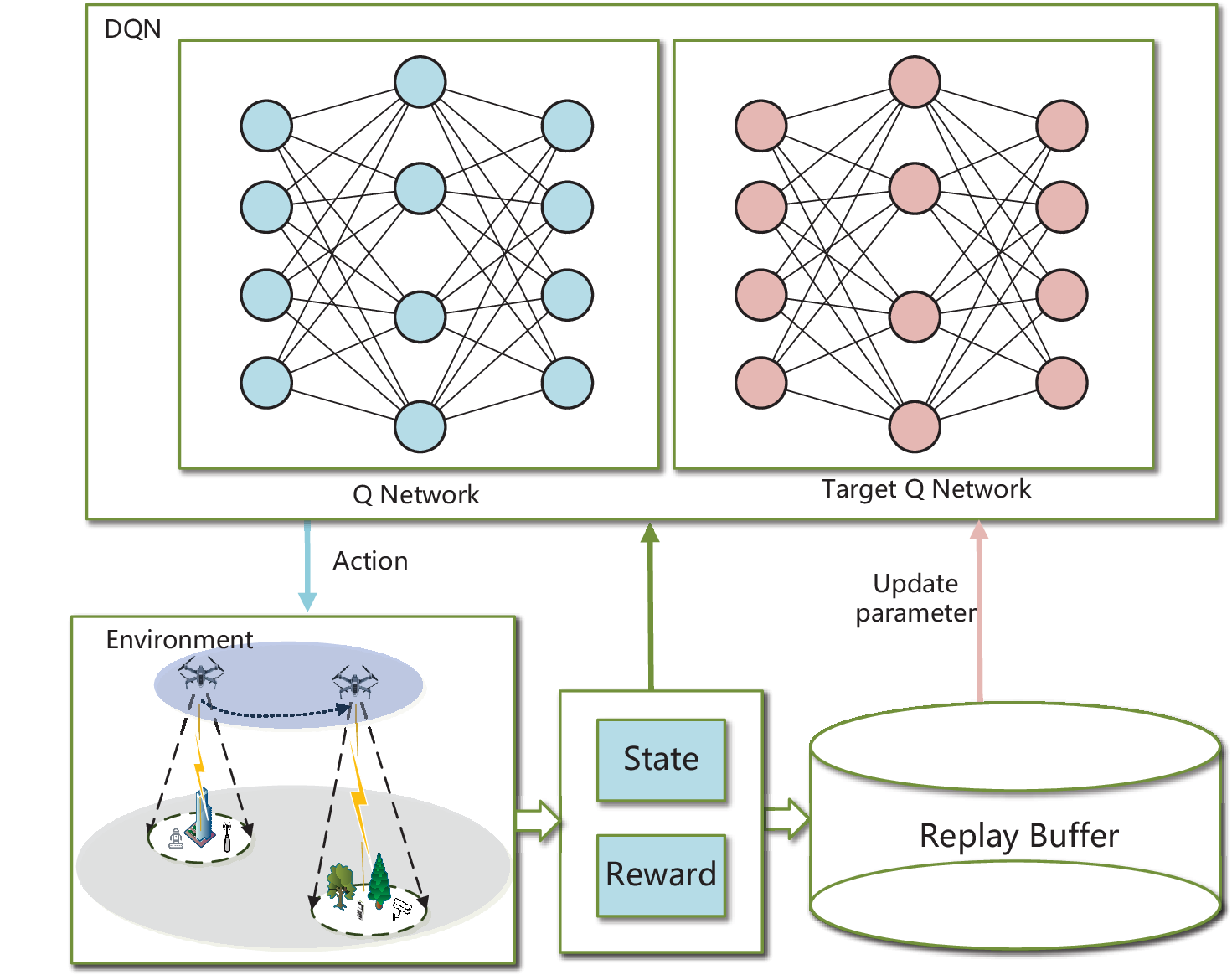}
\caption{The scenario of remote UAV control for data collection.}
\label{dqn}
\end{figure}

	Trajectory optimization problem is a sequential optimization problem where traditional iterative optimization algorithms have poor efficiency because it requires re-optimization when the trajectory distance changes. Deep Q-Network (DQN) offers a viable solution by combining the fitting capabilities of neural networks with the decision-making capabilities of reinforcement learning models, thereby efficient and real-time decision-making can be achieved. In our scenario, DQN allows using a single reinforcement learning network for the trajectory determination of varying distances. The overall process is illustrated in Fig. \ref{dqn}.

We formulate a Markov Decision Process(MDP) which consists of state space $\mathcal{S}$, action space $\mathcal{A}$ and reward function $R$. In our MDP, the UAV's dynamic model is presented in (\ref{cm}). We consider the UAV moving towards the target point along the shortest path. Therefore, trajectory determination can be considered as a one-dimensional acceleration decision problem. Consequently, the distance and velocity of the UAV can be expressed as
\begin{equation} \label{S1}
\begin{bmatrix}
d(k+1) \\
v(k+1)
\end{bmatrix}
=
\begin{bmatrix}
1 & -\delta \\
0 & 1
\end{bmatrix}
\begin{bmatrix}
d(k) \\ 
v(k)
\end{bmatrix}
+
\begin{bmatrix}
-\frac{1}{2} \delta^2 \\ 
\delta
\end{bmatrix}
a(k),
\end{equation}
where $d$ represents the distance of the UAV to the target position, $v$ represents the UAV's velocity, and $a$ denotes the UAV's acceleration. Besides, $d$, $v$ and $a$ are scalar here because the trajectory is a straight line. Due to the symmetry of energy consumption during acceleration and deceleration, we can decompose the optimization of UAV trajectory determination problem into the sum of an acceleration optimization stage and its symmetric problem. Consequently, our problem becomes determining how to reach half the required distance with minimal energy starting from an initial velocity of zero. This simplification allows us to remove the constraint of the final velocity to be zero which simplifies the reinforcement learning process because requiring the final velocity to be zero would necessitate penalty terms in the reward function. Moreover, by considering only the acceleration (or deceleration) process, the action space is halved. The following sections define the state space, action space, reward function, and algorithm.

	{\it 1) State space and Action space:} The state can be represented as ${\bf{s}}=[d;v]\in\mathcal{S}$. The action can be represented as $a\in\mathcal{A}$. In this MDP, the task of UAV trajectory determination involves determining the required acceleration for the next time slot based on the UAV's current distance to the target and its velocity, in order to achieve the next state.

	{\it 2) Reward function:} Our reward function comprises two sub-reward components. The first component, based on equation (\ref{CE}), penalizes the power consumed by the UAV during movement. The second component is an instantaneous reward granted when the UAV reaches the destination, encouraging it to fly towards the goal. The specific reward function $r = E_\text{f}(k) + R_2$, where $R_2$ represents the specific value of the reward for reaching the destination.

	{\it 3) Algorithm:} Based on the DQN algorithm and the UAV control model, our proposed algorithm is outlined in Algorithm \ref{alg:dqn}.	First, we initialize the replay buffer. Next, we initialize the Q-network $\mu_Q$ and the target Q-network $\mu_{Q'}$ with the same random network parameters. At the beginning of each training episode, we initialize the state of the UAV, which includes the UAV's distance and velocity. Actions are chosen with probability to explore the observation space; otherwise, the action with the highest Q-value is selected. The environment interacts with the chosen action, outputting the state {\bf{s}} and reward $R$ according to the formulas mention above.	We store the transition $({\bf{s}}, a, r, {\bf{s}}')$ into replay buffer and randomly sample a minibatch of transitions to train the network $\mu_Q$ using the Bellman equation. The target network $\mu_{Q'}$ is updated with $\mu_Q$ every $C$ steps. Each episode concludes when the UAV's distance reduces to zero.

\begin{algorithm} 
\caption{DQN-based Algorithm for giving reference trajectory} 
\label{alg:dqn} 
\begin{algorithmic}[1] 
\STATE Initialize replay buffer
\STATE Initialize action value network $\mu_Q$ and target action-value network $\mu_{Q'}$ with same random network parameters
\FOR{episode = 1 to $M$}
    \STATE Initialize state $\mathcal{S}$: initialize the distance and velocity of UAV 
    \WHILE{${\bf{s}}$ is not terminal}
        \STATE With probability to select a random action $a$
        \STATE otherwise select $a = \arg\max_a \mu_Q({\bf{S}}, a)$ 
        \STATE Execute action $a$ and observe next state ${\bf{s}}'$  and reward $r$
        \STATE Store transition $({\bf{s}}, a, r, {\bf{s}}')$ into replay buffer
        \STATE Sample random minibatch of transitions $({\bf{s}}_j, a_j, r_j, {\bf{s}}_j')$ from replay buffer
        \STATE Set $q_j = r_j + \gamma \max_{a} \mu_{Q'}({\bf{s}}'_j, a')$ if ${\bf{s}}'_j$ is not terminal
        \STATE $q_j = r_j$ if ${\bf{s}}'_j$ is terminal
        \STATE Use $q_j$ to train the action value network $\mu_Q$
        \IF{Every $C$ steps}
            \STATE Set $\mu_{Q'} = \mu_Q$
        \ENDIF
        \STATE Set ${\bf{s}} = {\bf{s}}'$
    \ENDWHILE
\ENDFOR
\end{algorithmic} 
\end{algorithm} 

\subsection{Communication Power Allocation for Data Uploading}
For given trajectory in Algorithm \ref{alg:dqn}, the flying time are fixed. The hovering time of each hovering points can be easily computed by (\ref{DR}) with maximum data rate for reducing the hovering energy consumption to improve the EE. As we mentioned above, the UAV is hovering for data collection, and flying for data uploading. Therefore, this optimization can be solved node by node without loss optimality. 
Using $i$-th flying process as the example,  we denote the flying time as $T_i$. 
Wuth given the variables $\boldsymbol\gamma$, and $\bf X$, the original problem for optimizing the power allocation to maximize the EE of UAV in $i$-th flying process can be reformulated as 
\begin{align}\label{p1}
\underset{{\bf P}(i)}{\max.}&\quad \frac{\sum\limits_{k=1}^{T_i}R_s(k)}{\sum\limits_{k=1}^{T_i}E_\text{d}(k)+A_1(i)} \\ \nonumber
 \text{s.t.}&\quad  \text{C2, C4},
\end{align}
where ${\bf P}(i)$ is the power allocation set of $i$-th flying process, and $A_1(i)=\sum\limits_{k=1}^{T_i}\left(E_\text{f}(k)+\gamma(k)E_\text{s}\right)$.  
For UAV-to-Satellite communication, the channel gain can be approximated only up to the height of satellite \cite{WYSJ}, indicating the data rates and transmitted power are symmetric across the time slots, denoted by $R(i)$ and $P(i)$, respectively. Therefore, the objective function in (\ref{p1}) can be rewritten as 
\begin{align}
f=\frac{D_\text{s}}{\frac{D_\text{s}P(i)}{R(i)}\delta+A_1(i)},
\end{align}
where the constraints are satisfied while maximizing $f$, which is to maximize the following part
\begin{align}
f_1=\frac{R(i)}{P(i)}=\frac{B\log_2\left(1+\frac{P(i)g_s}{\sigma_0^2}\right)}{P(i)}.
\end{align}
The first derivative of $f_1$ with respect to power is 
\begin{align}\label{FD}
\frac{\partial f_1}{\partial P(i)}=\frac{\log_2\exp\frac{g_s}{\sigma_0^2+P(i)g_s}-\log_2\left(1+\frac{P(i)g_s}{\sigma_0^2}\right)}{(P(i)^2)}.
\end{align}
It is easy to see the monotonicity decreasing of equation (\ref{FD}), which means concave property of $f$. Therefore, there exist a optimal power $P_\text{opt}$ to maximize the EE of UAV. The optimal power can be calculated by solving the following equation. 
\begin{align}
\log_2(\exp(\frac{g_s}{\sigma_0^2+P(i)g_s}))-\log_2\left(1+\frac{P(i)g_s}{\sigma_0^2}\right)=0.
\end{align}
It is evident that this equation has only one root, due to the monotonically increasing property of the left term.
To complete the data uploading during the flying time, there exist a minimum data rate $R_\text{min}(i)=\frac{D_s}{T_i}$. Then, we can obtain the minimum transmitted power by $P_\text{min}(i)=\left(2^{\frac{R_\text{min}(i)}{B}}-1\right)\cdot \frac{\sigma_0^2}{g_s}$. Since the objective function is concave, the optimal rate for each GD is reached when $P(i)=\max\left\{P_\text{opt}, P_\text{min}(i) \right\}$ due to the symmetry of data rate and transmitted power across all uploading time slots. To satisfy the constraint C4, the optimal transmitted power for all flying times slots is $P^o=\min\left\{P(i), P_\text{max}\right\}$.

{\it Remark 1:} When the $p_\text{max}\leq P_\text{min}(i)$, it indicates that the UAV cannot upload all data during the flight. In this situation, we allow the UAV to spend some time hovering at specific points. The new data collection process starts only after the previous uploading is completed.

%

\subsection{Sensing Scheduling for LQR Control}
Given the power allocation $\bf P$ and reference state $\bf X$, the EE depends only on sensing energy and flying energy. Therefore, the original optimization problem can be reformulated as
\begin{align}\label{p2}
\underset{\boldsymbol\gamma}{\min.}&\quad {\sum\limits_{k=1}^{T}E_{\text f}(k)+\gamma(k)E_{\text s}} \\ \nonumber
 \text{s.t.}&\quad  \text{C1,C7}.
\end{align}

In this optimization problem, the flying energy is related to the sensing scheduling. As given in equation (\ref{est}), the state information used for calculating the LQR output is determined by the current sensing scheduling decision.

To quantify the freshness of UAV state at the remote control center, we utilize the {\it age-of-information} (AoI) metric, $\phi(k)\in \mathbb{N}$, which is the time duration between previous successful reception time slot and current time slot $k$.   Accordingly, the updating rule for AoI can be expressed as
\begin{equation}
\phi(k+\Delta)=\left\{
{\begin{array}{cc}
                     \Delta & \lambda(k)=1, \\
                     \phi(k+\Delta-1)+1 & \text{otherwise}
                   \end{array}}
\right.
\end{equation}
Then, the estimation state in remote center for $\lambda(k)=1$ can be written as \cite{AOIIOT}
\begin{equation}
{\bf{\hat x}}(k+\Delta)=  {\bf A}^{\Delta}{\bf x}(k)+\sum\limits_{j=0}^{\Delta-1}{\bf A}^{j}{\bf B}\tilde{\bf u}(k+\Delta-j-1)+\tilde{\bf w}(k+\Delta-1),
\end{equation}
where $\tilde{\bf w}(k+\Delta-1)=\sum\limits_{j=1}^{\Delta}{\bf A}^{j-1}{\bf w}(k+\Delta-j)$, and $\tilde{\bf u}(k)={\bf{K}}(\hat{\bf x}(k)-{\bf x}_r(k+1))$ at the time slot $k$.  The predicated state in the remote center at time slot $k$ can be expressed as 
\begin{equation}
x(k+1)=  {\bf A}\tilde{\bf x}(k)+{\bf B}\tilde{\bf u}(k)+{\bf w}(k).
\end{equation}
\begin{figure}[h]
\centering
\includegraphics[width=1.05\columnwidth]{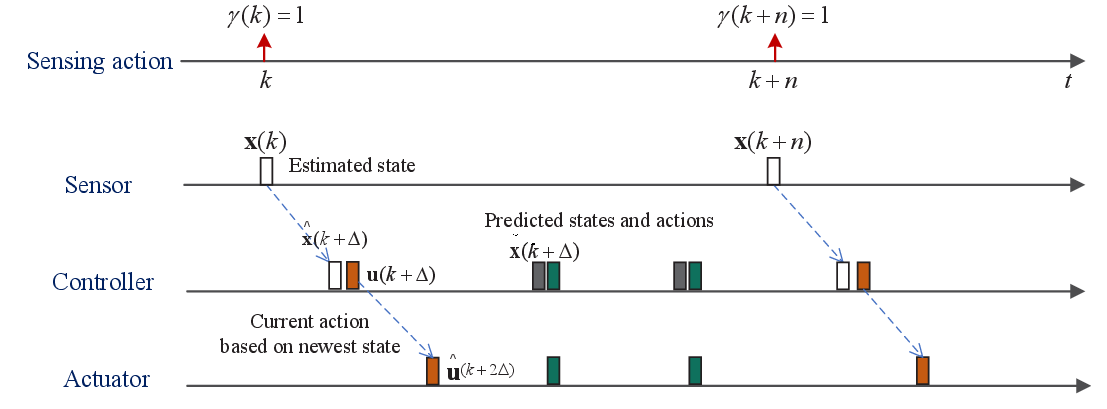}
\caption{The timing diagram of the remote UAV control system.  }
\label{time}
\end{figure}

Fig .\ref{time} illustrates the entire control process of the proposed architecture.  The first diagram showcases the sensing action, and the second diagram illustrates the sensing state at sensor. The third diagram shows the state  of UAV at remote control center for calculating the control commands, and the lastly diagram shows the received actions at the actuator. When $\gamma(k)=1$, the sensing operation is adopted to obtain the current UAV state ${\bf x}(k)$. Then, the state is transmitted to the remote control center. The controller computes the control commands based on the newest estimated state, and predict the state for further control commands calculation. After that, the actuator of UAV executes a set of actions for control in a period of time until receiving the new set of control commands. 

To guarantee the control stability in C7, we take the logarithm operation. Then, we can obtain the upper bound sensing interval of each time with 
\begin{equation}
q_\text{max}(k)= \frac{-\ln(1-\rho(k))}{\ln(\max \{\lambda_m\})}.
\end{equation}
As long as the scheduling duration between two sampling time instant is less than the upper bound, the control process is regarded as stable. 
  With the maximum sensing interval and the detailed control function, there exist a optimal sensing interval for each time slot. 
  Therefore, the problem (\ref{p2}) can be effectively solved by one-dimensional exhaustive search method.

\section{Simulation Results}

\begin{table}[h]
\centering
\caption{Main Simulation Parameters}\label{table1}
\begin{tabular}{|c|c|}
  \hline
  Communication Parameters & Value \\ \hline
  Carrier frequency &$f_c=2 $ [GHz]\\
  UAV-Satellite communication bandwith & $B_{\text s}=5$ [MHz]\\
  Ground-UAV communication bandwith & $B_0=0.5$ [MHz]\\
  Noise power & $\sigma_0^2=-110$ [dBm] \\
  The channel gain at 1 $\rm{m}$ & $\beta_0=-80$ [dB]\\
  Threshold of SNR & $\gamma_\text{th}=3$ [dB] \\
  Height of LEO & $H_s=1000$ [Km]\\
  Minimum elevation angle & $\varphi_\text{min}=50$ [${}^\circ$]\\
  Motion noise &$\sigma_x=\sigma_y=\sigma_z=0.005$ [m]\\
  State transmission noise & ${\bf R}=diag\{{\bf I}_{3\times3},0.1\cdot{\bf I}_{3\times3}\}$\\
  Weight matrix of control cost & $\boldsymbol{\epsilon}=0.5\cdot{\bf I}_{3\times3}$\\
  Weight matrix of LQR & ${\bf Q}={\bf I}_{6\times6}$ \\
  \hline
  Control Parameters & Value \\ \hline
  Time slot $T_s$ & $0.1$ [s] \\
  DQN State space & $\{[d;v]\}$ \\
  $d$&$d\in[0,250]$ [m] \\
  $v$&$v\in[0,50]$ [m/s]\\
  DQN action space & $\{0,1,\cdots,10\}$ [m/$\text s^2$]\\
  DQN reward for destination& $R_1=30000$ \\
  DQN target update frequency& $C=50$ \\
  \hline 
\end{tabular}
\end{table}

In this section, we assess the proposed method through extensive numerical simulations. The basic simulation setup is based on the existing specification, which is summarized in Table \ref{table1}. The performance of proposed method is investigated within $1000\text{m}\times1000 \text{m}\times 300\text{m}$ area, where 10 GDs and one UAV are deployed. 

\begin{figure}[h]
\centering
\includegraphics[width=1.05\columnwidth]{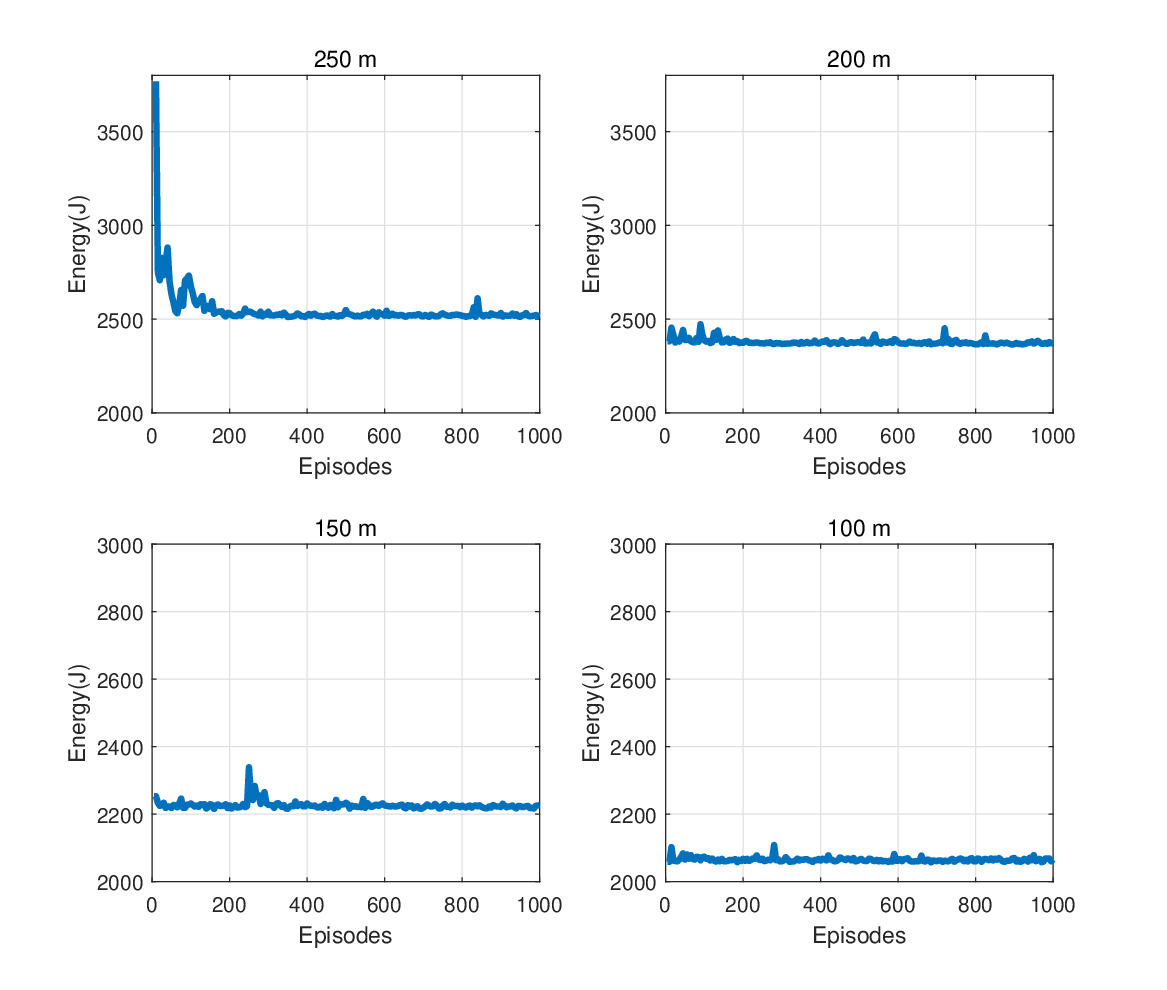}
\caption{The convergence of DQN-based trajectory determination.}
\label{convergence}
\end{figure}
\begin{figure}[h]
\centering
\includegraphics[width=1\columnwidth]{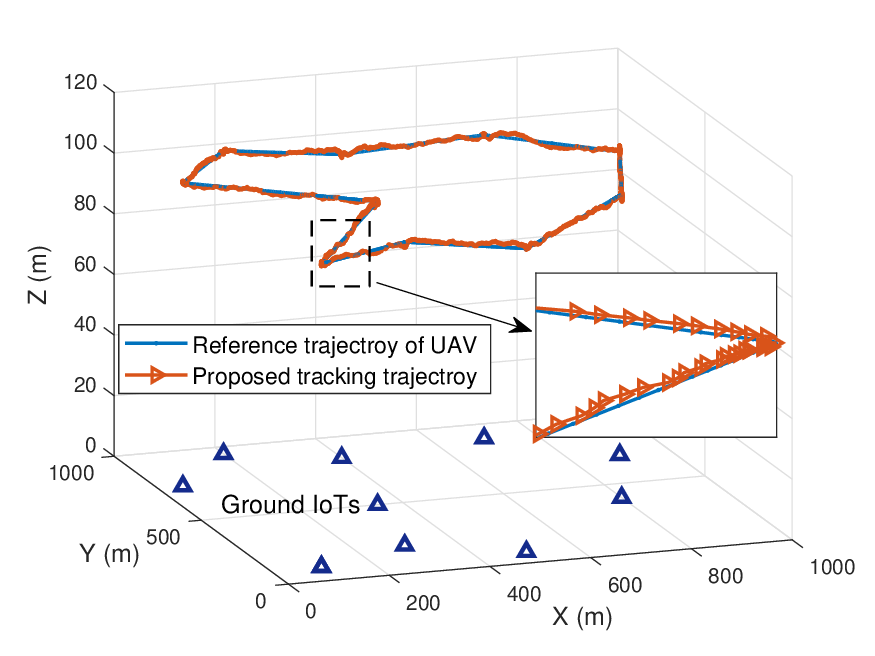}
\caption{The illustration our proposed remote control system for data collection.}
\label{trajectory}
\end{figure}

At the beginning, we delineate the trajectory based on the hover point, and then use DQN to determine the velocity and acceleration of the trajectory at the corresponding position, thereby providing reference trajectories for different time slots ${\bf{X}}_r$. In Fig. \ref{convergence}, we first depict the convergence performance of the DQN. During the training process, typical positions of the UAV relative to the target are selected at distances of $250\text{m}$, $200\text{m}$, $150\text{m}$, and $100\text{m}$. Initially, training commenced at a distance of $250\text{m}$. The first figure illustrates the rapid convergence of energy consumption in the trajectory of the UAV during this training phase. Subsequently, for the other training distances ($200\text{m}$, $150\text{m}$, and $100\text{m}$), due to the longer-distance trajectory encompassing the shorter-distance trajectories, fewer episodes are required to achieve a trajectory with low energy consumption. This training process demonstrates that our method exhibits good generalization capabilities when faced with trajectory of varying lengths.
\begin{figure}[h]
\centering
\includegraphics[width=1\columnwidth]{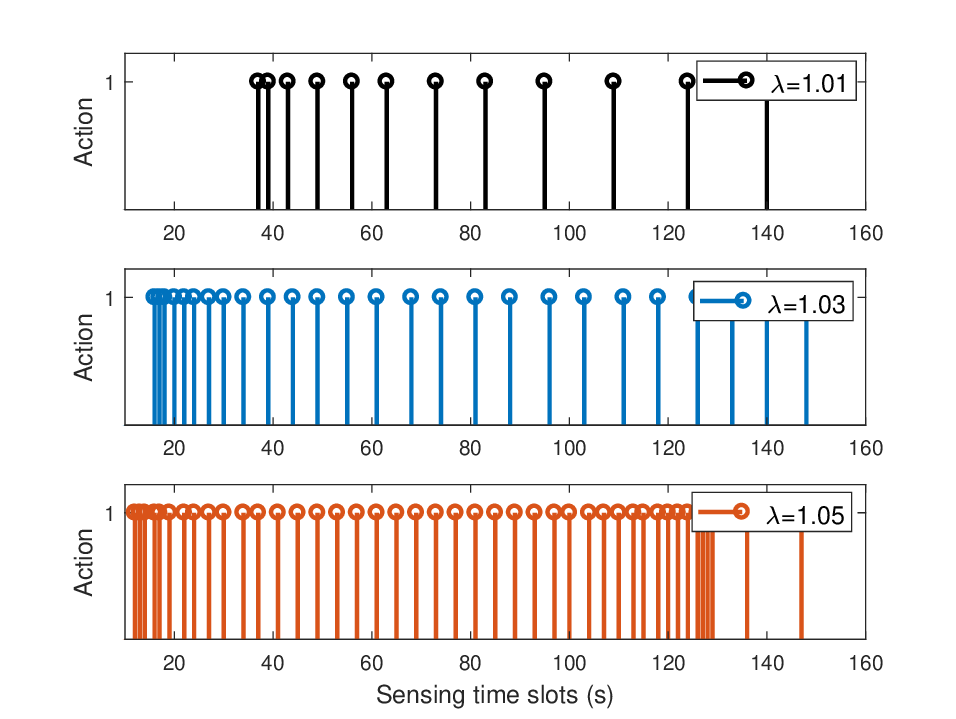}
\caption{The sensing scheduling under different control settings.}
\label{sensing}
\end{figure}
\begin{figure}[h]
\centering
\includegraphics[width=0.9\columnwidth]{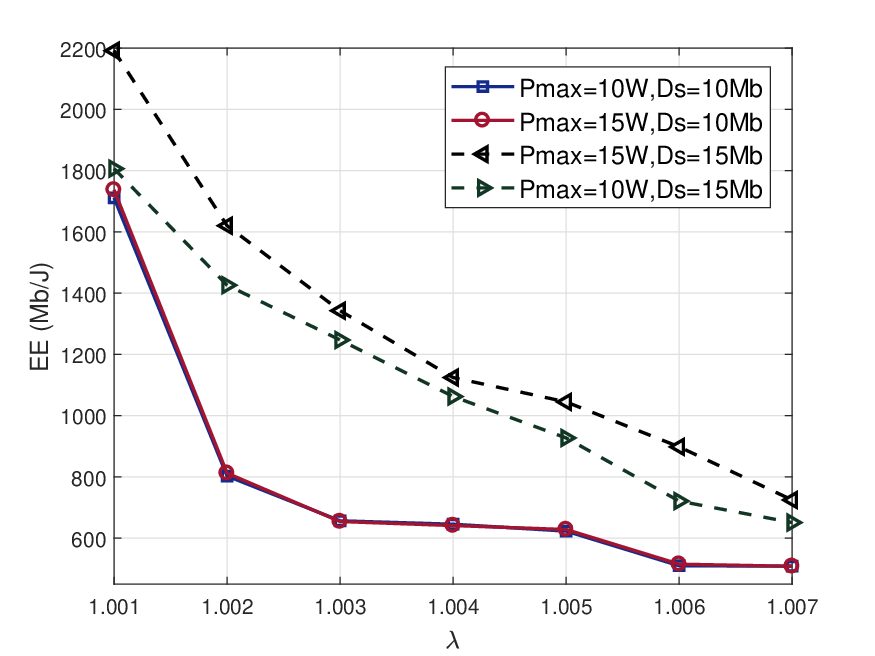}
\caption{The performance of EE vs. different $\lambda$.}
\label{EEvslambda}
\end{figure}

Next, we illustrate the trajectory of our proposed method. As shown in Fig .\ref{trajectory}, the proposed trajectory tracking method can obtain a great tracking performance with the reference state. The sub-figure showcase the detailed UAV motions.

Fig. \ref{sensing} illustrates the sensing scheduling actions under various system settings. As mentioned in Section \ref{sec:II}, a higher eigenvalue $\lambda=\max(\lambda_m)$ of the state transition matrix results in greater instability of the control system. Consequently, this necessitates an increased sensing frequency to ensure control stability. Notably, sensing operations decrease around the $140\text{s}$, when the UAV is hovering to execute data collection. This indicates a high probability of successful sensing and a stable state during UAV hovering. Therefore, sensing operations should cease in this situation to reduce energy consumption.
\begin{figure}[h]
\centering
\includegraphics[width=0.9\columnwidth]{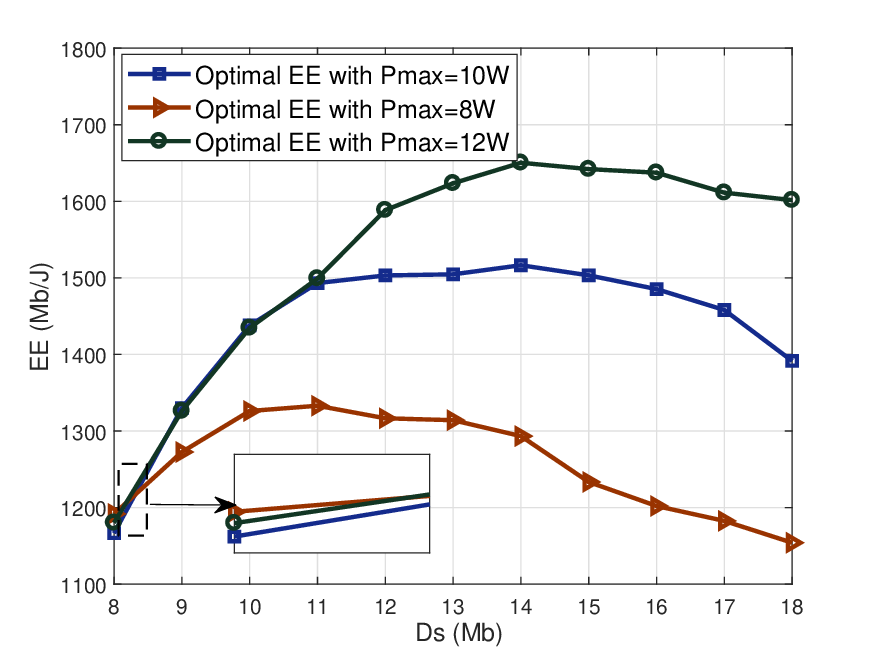}
\caption{The performance of EE vs. the minimum data size.}
\label{EEvsD}
\end{figure}
\begin{figure}[h]
\centering
\includegraphics[width=0.9\columnwidth]{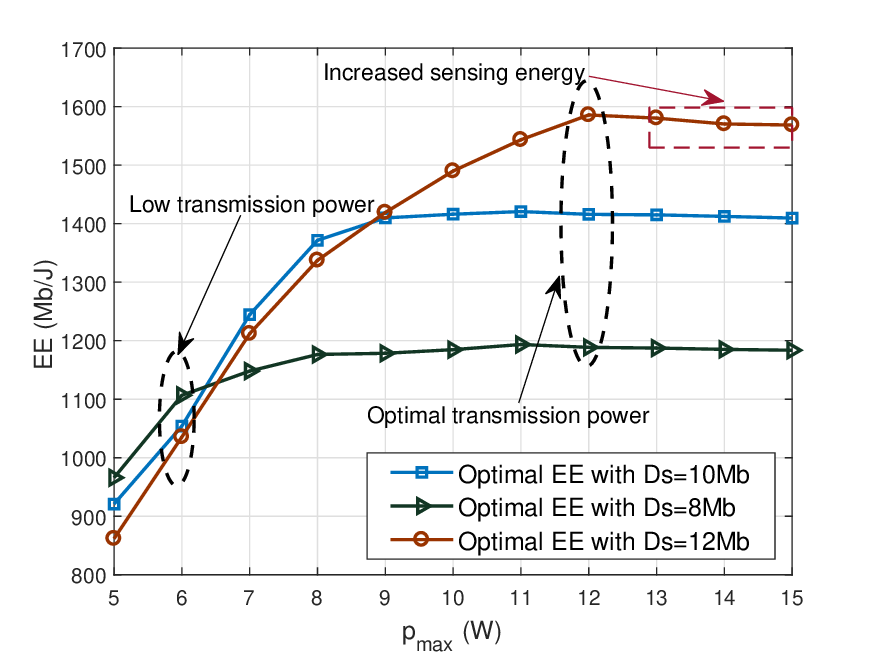}
\caption{The performance of EE vs. the max transmitted power.}
\label{EEvsP}
\end{figure}

In Fig. \ref{EEvslambda}, we investigate the EE under different control settings $\lambda$. An increase in $\lambda$ leads to greater action energy consumption by the UAV to counter control instability during state transitions. Given the minimum data collection size as $D_\text{s}=10\rm{Mb}$, EE is comparable between $p_\text{max}=10\rm{W}$ and $p_\text{max}=15\rm{W}$, as the optimal power allocation derived in Section \ref{sec:III} is below $p_\text{max}$. Given the minimum data size $D_\text{s}=15\rm{Mb}$, the performance of EE with $p_\text{max}=15\rm{W}$ is better than $p_\text{max}=10\rm{W}$, which means the optimal power is greater than $p_\text{max}$. The higher power accelerates the data uploading process and reduces the UAV's flying time. The EE for $D_\text{s}=15\rm{Mb}$ is much greater than $D_\text{s}=10\rm{Mb}$, as the UAV's flying time is not fully utilized for smaller data sizes.

To reveal the relationship between maximum transmission power and minimum data collection size, we conduct the simulation in Fig. \ref{EEvsD}. For a certain $p_\text{max}$,  the performance of EE increases with $D_\text{s}$ when the data size is small, which also agrees the observations in Fig. \ref{EEvslambda}. Keep increasing the data size, the EE decreases due to the increased hovering and uploading duration. For the large data size,  the EE is up to the maximum power, by reducing the time duration for the data collection task. For relative small data size, the EE depends on optimal power not the maximum power.

In Fig. \ref{EEvsP}, we can further explain the result in Fig. \ref{EEvsD}. The EE with small data size outperforms than relative large data size at the low transmission power region. The small data size leads faster data uploading time, which can be done during the flying process. Increased data size in low transmission power will increase the extra hovering time for  data uploading time, resulting in worse EE. At the high transmission power region,  the uploading time of small data size is less than the flying time, causing the insufficient utilization of UAV flying and low EE.  There exists a optimal transmission power the each data size. Keep increasing the transmission, the EE will decrease due to the increased sensing energy, which reflect there is a basic trade-off between communication, UAV flying and sensing.

\section{Conclusion}
In this paper, we proposed a Satellite-UAV architecture for data collection, where the UAV is controlled by a remote control center through satellite links. We first established a framework for UAV control, communication, and state sensing. A joint UAV trajectory, communication power, and sensing strategies optimization problem is presented to maximize the EE of UAV, while ensuring the control stability and communication reliability. The numerical results consistently demonstrate that our proposed method can achieve a great performance on EE without compromising control stability. The simulation results provide comprehensive designing strategies with different data size and power limitations.
Furthermore, our method can effectively balance a good trade off between UAV control, communication, and state sensing.
In the future the co-design method in multi-UAV scenarios can be another interesting topic.


\bibliographystyle{IEEETran}
\bibliography{ref}

\end{document}